\documentclass[usenatbib,usegraphicx]{mn2e}
\usepackage{float}
\usepackage{color}
\usepackage{amsmath}
\usepackage{enumerate}

\def\araa{ARA\&A}%
\def\apj{ApJ}%
\def\apjs{ApJS}%
\def\aap{A\&A}%
\def\mnras{MNRAS}%
\title{Magnetohydrodynamics on an unstructured moving grid}
\author[R.~Pakmor, A.~Bauer \& V.~Springel]
       {Ruediger~Pakmor$^1$, Andreas~Bauer$^1$ \& Volker~Springel$^{1,2}$\\
$^1$Heidelberger Institut f\"{u}r Theoretische Studien,
        Schloss-Wolfsbrunnenweg 35, 69118 Heidelberg, Germany\\
$^2$Zentrum f\"ur Astronomie der Universit\"at Heidelberg, Astronomisches
Recheninstitut, M\"{o}nchhofstr. 12-14, 69120 Heidelberg, Germany}

\date{Accepted 2011 August 04. Received 2011 August 01; in original form 2011 June 06 }

\begin{document}
  \maketitle

  \label{firstpage}

  \begin{abstract}
    Magnetic fields play an important role in astrophysics on a wide
    variety of scales, ranging from the Sun and compact objects to
    galaxies and galaxy clusters.  Here we discuss a novel
    implementation of ideal magnetohydrodynamics (MHD) in the moving
    mesh code {\small AREPO} which combines many of the advantages of
    Eulerian and Lagrangian methods in a single computational
    technique. The employed grid is defined as the Voronoi
    tessellation of a set of mesh-generating points which can move
    along with the flow, yielding an automatic adaptivity of the mesh and a
    substantial reduction of advection errors.  Our scheme solves
    the MHD Riemann problem in the rest frame of the Voronoi
    interfaces using the HLLD Riemann solver. To satisfy the
    divergence constraint of the magnetic field in multiple
    dimensions, the Dedner divergence cleaning method is applied. In a
    set of standard test problems we show that the new code produces
    accurate results, and that the divergence of the magnetic field is
    kept sufficiently small to closely preserve the correct physical
    solution. We also apply the code to two first application
    problems, namely supersonic MHD turbulence and the spherical
    collapse of a magnetized cloud. We verify that the code is able to
    handle both problems well, demonstrating the applicability of this
    MHD version of {\small AREPO} to a wide range of problems in
    astrophysics.
  \end{abstract}


  \section{Introduction}

  In many astrophysical systems, the gas is partially or fully
  ionized, such that its conductivity is very high while its internal
  viscosity is very low, implying that it can be treated in the limit
  of ideal magnetohydrodynamics. This is for example the case for the
  diffuse gas found in clusters of galaxies or in the halos of
  ordinary galaxies, and in most of the volume of the interstellar
  medium, where magnetic fields probably play an important role in
  regulating star formation. Also, magnetic fields are thought to be a
  key ingredient in mediating angular momentum in accretion disks
  through the magnetorotational instability
  \citep{Balbus1998}. Magnetic fields are also critical for launching
  relativistic outflows in some compact objects, and likely have a
  major influence on core-collapse supernovae.

  There is hence ample motivation to outfit numerical codes for
  hydrodynamics with an additional treatment of magnetic fields, 
  prompting the development of a large number of such codes in astrophysics
  over the years
  \citep[e.g.][]{Stone1992,Ziegler1997,Brandenburg2002,Fromang2006,Rosswog2007,stone2008a,Dolag2009,Collins2010}.
  Unfortunately, an important technical complication makes this far
  from straightforward. While the continuum equations of ideal MHD
  preserve $\nabla\cdot \textbf{B}= 0$ in an initially divergence free
  field, this is not necessarily the case for discretized versions of the
  equations. Here the relevant difference operators will tend to pick
  up or produce a non-vanishing value for $\nabla\cdot \textbf{B}$,
  and once such a spurious `magnetic monopole' has been produced, it
  has the tendency to become quickly larger in any non-trivial MHD flow,
  rapidly rendering the calculated solution unphysical. Simply
  ignoring this problem is not a viable strategy, except for a
  small class of very simple and well behaved test problems.

  Much work has therefore been done on developing discretization
  schemes for MHD that circumvent this problem. A detailed comparison
  of different approaches can be found in \citet{toth2000a}. All approaches
  either try to construct the discretization such that it inherently forces
  $\nabla\cdot \textbf{B}$ to be zero or modify the equations of ideal MHD
  to keep it small.
  
  A straight-forward way to get rid of $\nabla\cdot \textbf{B}$ field components has been
  proposed by \citet{Brackbill1980}. By projecting out the $\nabla\cdot \textbf{B}$
  component through a Helmholtz decomposition or a similar field cleaning
  technique one can regain a divergence-free magnetic field at any time.
  
  Another approach is to add additional terms that diffuse away the error once
  it appears \citep[see, e.g.][]{powell1999a,dedner2002a,keppens2003a}. In particular
  the so-called Dedner cleaning technique \citep{dedner2002a} has
  proven to be quite robust and in most cases yields results that are
  of comparable accuracy as those obtained with the ``constrained
  transport'' scheme of \citet{evans1988a} which fulfills the divergence
  constraint by construction. Here the transport of magnetic flux at the discretized
  level of the equations is carefully formulated in terms of loop
  integrals over electric fields at the mesh-edges, such that the magnetic
  flux and the $\nabla\cdot \textbf{B}=0$ constraint are conserved to
  machine precision.

  Unfortunately, the constrained transport method is only easily
  tractable for Cartesian meshes, and it is presently unclear whether
  it can be adapted to dynamic unstructured meshes such as the one we
  consider in this paper. We shall hence resort to the Dedner cleaning
  technique. We note that yet another possibility for working around
  the $\nabla\cdot \textbf{B}$ problems lies in using the
  vector potential  \citep[e.g.][]{Price2010b}, or the
  so-called Euler potentials
  \citep[e.g.][]{Rosswog2007,Dolag2009}. While both of these
  approaches formally guarantee $\nabla\cdot \textbf{B}=0$, they are
  met with a number of serious problems in practical applications, in
  particular with respect to correctly accounting for magnetic
  dissipation in turbulent flows \citep{Brandenburg2010}, hence we do
  not consider them here.

  The {\small AREPO} code introduced by \citet{springel2010a}
  represents a novel type of astrophysical simulation code, taking
  on an intermediate role between traditional Eulerian mesh-based
  hydrodynamics and mesh-free smoothed particle hydrodynamics. In
  {\small AREPO}, a Voronoi tessellation is employed to construct an
  unstructured mesh, which is then used to solve the equations of
  ideal hydrodynamics with a second-order accurate finite volume
  scheme based on Godunov's method. A Cartesian mesh is in fact a
  special case of a Voronoi tessellation, and here {\small AREPO}'s
  basic fluid dynamical approach is equivalent to the widely employed
  MUSCL-Hancock scheme \citep{Leer1984,Toro1997}. However, one crucial
  difference is that in {\small AREPO} the mesh-generating points can
  be moved arbitrarily. In particular, they can be moved with the
  local flow velocity such that the mesh follows the motion of the
  gas, providing quasi-Lagrangian behaviour. In this mode, advection
  errors are greatly reduced compared to ordinary Eulerian techniques
  while at the same time their accuracy for shock waves and fluid instabilities is
  retained. This approach is then also not limited by global timestep
  contraints coming from the largest velocity in a system; instead,
  only the local timestep constraints apply, making the scheme ideal for
  systems with large bulk velocities, such as accretion disks, or
  galaxy collisions. 

  We note that these attractive features have already led other groups
  to start developing new codes based on similar design
  principles. For example, {\small TESS} by \citet{Duffell2011} is a
  new moving-mesh code that includes an extension to relativistic
  hydrodynamics and magnetic fields, albeit so far only in 2D, in
  serial, and without a control of the $\nabla\cdot\textbf{B}$
  error.

  In this study, we present the details of our new implementation of
  MHD in the massively parallel {\small AREPO} code. We focus on a
  concise description of the numerical implementation of the MHD part
  of the code, and a discussion of a set of basic test problems. A
  detailed account of the mesh construction algorithms and
  parallelization strategies can be found in \citet{springel2010a}.

  The paper is structured as follows. Section~\ref{sec:implementation}
  describes our implementation of MHD in the {\small AREPO} code. It
  is followed by a discussion of several standard test problems for
  MHD in Section~\ref{sec:tests}. In Section~\ref{sec:turbulence} we
  show that the code can be used to simulate MHD
  turbulence. Additionally we apply the code to an important
  prototypical astrophysical application, the collapse of a magnetized
  cloud, in Section~\ref{sec:collapse}. Finally we give our
  conclusions in Section~\ref{sec:conclusions}.

   \section{Implementation}
 
   \label{sec:implementation}
    \subsection{The equations of magnetohydrodynamics}
    The equations of ideal magneto-hydrodynamics can be written as a
    system of conservation laws,
    \begin{equation}
      \frac{\partial \textbf{U}}{\partial t} + \nabla\cdot \textbf{F}
      = 0 ,
    \end{equation}
    for a vector of conserved variables $\textbf{U}$ and a flux function $\textbf{F}(\textbf{U})$, which are given in the local
    restframe by
    \begin{equation}
      \textbf{U} = \left( \begin{array}{c} \rho \\ \rho \textbf{v} \\ \rho e \\ \textbf{B} \end{array} \right)
      \ \ \ \ \ \ \
      \textbf{F}(\textbf{U}) = \left( \begin{array}{c} 
        \rho \textbf{v} \\
        \rho \textbf{v} \textbf{v}^T + p - \textbf{B} \textbf{B}^T \\
        \rho e \textbf{v} + p \textbf{v} - \textbf{B} \left( \textbf{v} \cdot \textbf{B} \right) \\
        \textbf{B} \textbf{v}^T - \textbf{v} \textbf{B}^T
      \end{array} \right).
    \end{equation}
    Here, $p = p_{\rm gas} + \frac{1}{2} \textbf{B}^2$ is the total
    gas pressure and
    $e = u + \frac{1}{2} \textbf{v}^2 + \frac{1}{2 \rho} \textbf{B}^2$
    is the total energy per unit mass, with $u$ denoting the thermal
    energy per unit mass. $\rho$, $\textbf{v}$ and $\textbf{B}$ give the
    local gas density, velocity and magnetic field strength, respectively. 

    For $\textbf{B}=0$, these equations reduce to ideal hydrodynamics,
    which is treated by our version of {\small AREPO} in the same
    manner as described by \citet{springel2010a}, in particular with
    respect to the technical details of Voronoi mesh construction,
    gradient estimation, and parallelization. In the interest of
    brevity, we will therefore restrict ourselves in the following to a
    description of the aspects relevant for the new MHD implementation.

   \subsection{Solving the Riemann problem}

   To estimate the fluxes over an interface we follow Godunov's
   approach and solve an approximate Riemann problem normal to the
   interface.  We obtain the initial left and right state at the
   interface by spatially 
   extrapolating the primitive variables at the centres of both
   neighboring cells to the mid-point of the interface, and
   by predicting them half a timestep
   forward in time. While the time extrapolation is done as described in
   \citet{springel2010a}, for the extrapolation in space we use a refined
   approach based on  \citet{darwish2003a}. To extrapolate a
   primitive variable from the left cell to the interface, according
   to their proposal one first
   defines the scalar $r$ by
    \begin{equation}
      r_{\rm L} = \frac{2 \nabla \phi_{\rm L} \cdot \textbf{r}_{\rm
          LR}}{\phi_{\rm R} - \phi_{\rm L}} - 1,
    \end{equation}
    where $\phi_{\rm L,R}$ are the values of the primitive variable $\phi$
    at the centers of the cells left and right from the interface,
    $\textbf{r}_{\rm LR}$ is the vector from the center of the left cell to
    the center of the right cell and $\nabla \phi_{\rm L}$ the gradient of
    $\phi$ at the center of the left cell.  The value $\phi_{\rm
      L}^{\star} $ of $\phi$ on the
    interface extrapolated from the left cell can then be calculated
    as
    \begin{equation}
      \phi_{\rm L}^{\star} = \phi_{\rm L} + \frac{1}{2} \Psi \left( r_{\rm L} \right)
      \left( \phi_{\rm R} - \phi_{\rm L} \right),
    \end{equation}
    where $\Psi \left( r_{\rm L} \right)$ is the slope limiting
    function.  The extrapolation from the right cell to the interface
    is done likewise.  Our usual choice for $\Psi$ is the van-Leer
    limiter \citep{vanleer1974a}, defined by
    \begin{equation}
      \Psi \left( r \right) = \frac{ r + \left| r \right| }{ 1 + \left| r \right| },
    \end{equation}
but other choices are in principle possible as well. Note that 
this slope limiting procedure recovers the common formulation for a
Cartesian grid as a special case.

To solve the Riemann problem in the MHD case, we use a three-step
approach that is based on the use of up to three different Riemann solvers in the following
order:
    \begin{enumerate}[1)]
     \item The HLLD solver \citep{miyoshi2005a}
     \item The HLL solver \citep{harten1983a}
     \item The Rusanov solver \citep{rusanov1961a}
    \end{enumerate}

    Each of these Riemann solvers returns fluxes over the interface as
    well as the state of the primitive variables at the interface.
    Starting with the HLLD solver, we check whether the pressure $p^*$
    of the solution at the interface is positive. If this is not the
    case, the solution is not valid and is discarded. Instead, we
    revert to the HLL solver and try again. If $p^*$ is still
    negative, we also discard this solution and finally employ the
    Rusanov solver which guarantees a valid solution. We found this
    approach to provide a good compromise in terms of accuracy and
    robustness.
    
    \subsection{The divergence constraint}
    
In addition to satisfying the MHD equations that govern the time
evolution of the magnetic field, 
the field always has to obey the constraint
    \begin{equation}
      \nabla \cdot \textbf{B} = 0.
    \end{equation}
    It can be shown that the differential form of the MHD evolution equations keep a
    magnetic field divergence free if it has been in this state
    initially. Numerically, however, due to discretization errors this
    is not the case in general. In one dimension, the constraint reduces
    to ${\partial B_x} / {\partial x} = 0$, which is equivalent to
    requiring a constant magnetic field in the $x$-direction, and is
    hence easily fulfilled. In two- or three dimensions, however, it
    is in general non-trivial to satisfy the divergence constraint. 

    Today, the most widely used approach to ensure $\nabla \cdot
    \textbf{B} = 0$ in numerical MHD is the constrained transport
    scheme \citep{evans1988a,gardiner2005a} which guarantees that the
    divergence of the magnetic field remains zero by
    construction. However, it is very difficult, if at all possible, to
    adapt this scheme to an unstructured moving grid.  Therefore we
    here use an alternative approach proposed by \citet{dedner2002a},
    with the goal to keep the divergence of the magnetic field always
    sufficiently small rather than guaranteeing it to be exactly
    zero. This can be viewed as the next best alternative compared to
    constrained transport.

   We apply the GLM-MHD approach \citep{dedner2002a}
    by introducing an additional (conserved) scalar $\psi$ which is related to the
    divergence of the magnetic field. The new vector of conserved
    quantities and the new flux function include $\psi$, and are given by
    \begin{equation}
      \textbf{U} = \left( \begin{array}{c} \rho \\ \rho \textbf{v} \\ \rho e \\ \textbf{B} \\ \psi \end{array} \right)
      \ \ \ \ \ \ \
      \textbf{F}(\textbf{U}) = \left( \begin{array}{c} 
        \rho \textbf{v} \\
        \rho \textbf{v} \textbf{v}^T + p - \textbf{B} \textbf{B}^T \\
        \rho e \textbf{v} + p \textbf{v} - \textbf{B} \left( \textbf{v} \cdot \textbf{B} \right) \\
        \textbf{B} \textbf{v}^T - \textbf{v} \textbf{B}^T + \psi I \\
        c_h^2 \textbf{B}
      \end{array} \right).
    \end{equation}
    Here $c_h$ is a positive constant. In addition, we adopt  a source term leading to an exponential decay of $\psi$,
    \begin{equation}
      \frac{\partial \psi}{\partial t} = - \frac{c_h^2}{c_p^2} \psi,
    \end{equation}
    where a second constant $c_p$ has been added. 
   We include this source term in the time integration using operator splitting. As shown in \citet{dedner2002a}, it is possible
    to solve the equations for the fluxes of $B_x$ and $\psi$
    separately, and to use an ordinary MHD Riemann solver in a second step applying the
    value of $B_x$ at the interface that is obtained in the first
    step. The fluxes are then formally given by
    \begin{equation}
      \label{eq:fluxes:divb}
      \textbf{F} \left( \begin{array}{c} B_x \\ \Psi  \end{array} \right) = \left( \begin{array}{c} \Psi^* \\ c_h^2 B_x^* \end{array} \right).
    \end{equation}

    There are no physical constraints on the choice of the two
    constants $c_h$ and $c_p$ but they have to be chosen such that the
    scheme is stable and prevents the divergence of the magnetic field
    from becoming so large that the solution of the problem is
    affected. For most standard test cases, both goals can be
    fulfilled without much of a problem. For turbulent motions,
    however, this is more challenging. In particular, the choice of $c_h$,
    which acts as the velocity with which the divergence of the magnetic field is
    advected away, requires some care.  If it is too small, the
    transport will not be efficient enough and the scheme can become
    unstable due to a local build up of a  large divergence error in the magnetic
    field. Conversely, if it is set too large, the physical
    solution of the problem can be impacted.
    
    A further difficulty is that the {\small AREPO} code evolves all
    cells on individual timesteps according to their local time-stepping
    criterion, as described by \citet{springel2010a}. 
    Equation~(\ref{eq:fluxes:divb}) directly
    shows that $c_h$ must be chosen globally, because a different
    $c_h$ at two opposite interfaces of a cell would lead to a net
    flux of $\Psi$ even for a constant homogenous magnetic field. The
    same argument demands that we may change $c_h$ only when all cells
    are synchronized. As we do not know at that point in time how large it
    needs to be to keep the scheme stable until the next update opportunity, we
    assign the maximum signal velocity of all cells to $c_h$. Assuming
    that our grid motion is close to Lagrangian, the maximum signal
    velocity is given by the fastest magneto-acoustic wave
    \begin{equation}
      c_h = \mathrm{max}_i \left( c_{f} \right).
    \end{equation}
    Since $c_h$ acts as a velocity and will be larger than the local
    signal velocity for many cells, we need to introduce an additional
    limitation to the timestep of a cell, similar to the
    Courant-criterion, to keep the ordinary algorithm stable:
    \begin{equation}
      \Delta t_i < C_{\mathrm{CFL}} \frac{r_i}{c_h} .
    \end{equation}
    Here, $C_{\mathrm{CFL}}$ is the same constant used for limiting
    the hydrodynamical timestep and $r_i$ is the effective radius of
    the cell.
    The choice of $c_p$ turns out to be less problematic. As proposed
    by \citet{dedner2002a}, we simply adopt
    \begin{equation}
      c_p = \sqrt{ 0.18\ c_h }.
    \end{equation}

    \begin{figure*}
      \centering
      \includegraphics{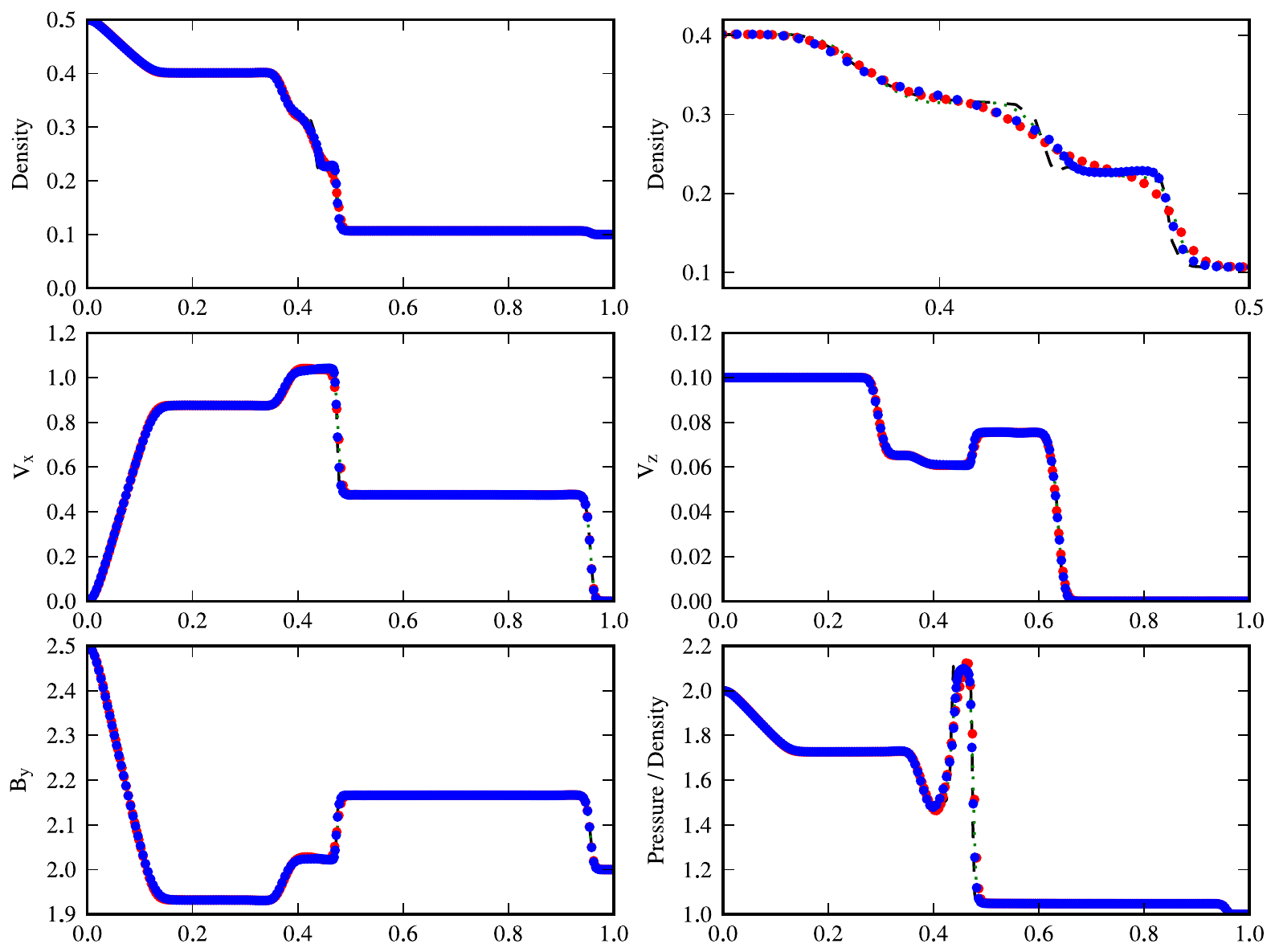}
      \caption[Shock-Tube]
      {Results of a one-dimensional shock-tube described
        by \citet{keppens2004a} with $\gamma =
        {5}/{3}$. The left and right initial states are given by
        $(\rho, p, v_x, v_y, v_z, B_x, B_y, B_z ) = (0.5, 1.0, 0, 1.0,
        0.1, 1.0, 2.5, 0)$ and $(0.1, 0, 0, 0, 0, 1.0, 2.0, 0)$,
        respectively. Shown are numerical solutions at $t = 0.08$
        for a resolution of $250$ cells for $\rho$ (top row, top right panel zooms
        into the density discontinuity), $v_x$ (middle left),
        $v_z$ (middle right), $B_y$ (bottom left), and $p/\rho$
        (bottom right). The plots show results for runs with a static/moving grid
        using the HLLD Riemann solver (green dotted/black dashed line) and the
        Lax-Friedrich solver (red/blue dots). The position of the dots is given by
        the centers of the cells.}
      \label{fig:shock}
    \end{figure*}

  \subsection{MHD on a moving grid}

  The fluxes described above are valid only for a static grid. If the
  grid itself moves, additional terms will be needed to account for
  the movement. The flux over an interface moving with velocity
  $\textbf{w}$ can be described as a combination of the flux over a
  static interface and an advection step owing to the movement of the
  interface:
    \begin{equation} \begin{aligned}
      \textbf{F}_m(\textbf{U}) &= \textbf{F}_s(\textbf{U}) - \textbf{U} \textbf{w}^T = \\
      &=
      \left( \begin{array}{c} 
        \rho \textbf{v} \\
        \rho \textbf{v} \textbf{v}^T + p - \textbf{B} \textbf{B}^T \\
        \rho e \textbf{v} + p \textbf{v} - \textbf{B} \left( \textbf{v} \cdot \textbf{B} \right) \\
        \textbf{B} \textbf{v}^T - \textbf{v} \textbf{B}^T + \psi I \\
        c_h^2 \textbf{B}
      \end{array} \right) -
      \left( \begin{array}{c} 
        \rho \textbf{w} \\
        \rho \textbf{v} \textbf{w}^T \\
        \rho e \textbf{w} \\
        \textbf{B} \textbf{w}^T \\
        \psi \textbf{w} \\
      \end{array} \right).
  \end{aligned} \end{equation} However, the straightforward approach
to use the approximate solution of the Riemann problem in the
rest-frame, $\textbf{F}_s$, and then to advect the state on the interface
with the velocity of the interface, turns out to be unstable. This is
mainly a result of our usual choice of the grid velocities, which
cause the velocity of the interface to be very close to the fluid
velocity at the interface, $\textbf{w} \approx \textbf{v}$, and the
mass flux over the moving interface is accordingly close to zero. As
the mass flux over the interface in the rest frame is calculated from
an approximative Riemann solver, there is always a small error. And because the
mass flux in the moving frame is very small, this error can cause a
sign change, destroying the upwind property of the scheme and
making it unstable.  This behavior can be avoided by calculating the
fluxes in the rest-frame of the interface. There the new vectors of
conserved variables and fluxes are
    \begin{equation}
      \textbf{U}^\prime = \left( \begin{array}{c} \rho \\ \rho \left( \textbf{v}-\textbf{w} \right) \\
      \rho e^\prime \\ 
      \textbf{B} \\ \psi \end{array} \right), 
    \end{equation}
   \begin{equation} \begin{aligned}
      \textbf{F}^\prime(\textbf{U}^\prime) &=
      \left( \begin{array}{c}
        \rho \left( \textbf{v} - \textbf{w} \right) \\
        \rho \left( \textbf{v} - \textbf{w} \right) \left( \textbf{v} - \textbf{w} \right)^T + p - \textbf{B} \textbf{B}^T \\
        \rho e^\prime \left( \textbf{v} - \textbf{w} \right) + p \left( \textbf{v} - \textbf{w} \right) - \textbf{B} \left( \left(\textbf{v}-\textbf{w}\right) \cdot \textbf{B} \right) \\
        \textbf{B} \left( \textbf{v} - \textbf{w} \right)^T - \left( \textbf{v} - \textbf{w} \right) \textbf{B}^T + \psi I \\
        c_h^2 \textbf{B}
      \end{array} \right) \\
      &= \left( \begin{array}{c}
        Q_1 \\ \textbf{Q}_2 \\ Q_3 \\ \textbf{Q}_4 \\ Q_5
      \end{array} \right),
    \end{aligned} \end{equation}
    with $e^\prime = e - \frac{1}{2} \textbf{v}^2 + \frac{1}{2} \left(\textbf{v}-\textbf{w}\right)^2$.
    To get the fluxes in the rest frame of the mesh from the fluxes in the moving frame we have to add some additional terms:
    \begin{equation} \begin{aligned}
      &\textbf{F}_m(\textbf{U}) = \textbf{F}_s(\textbf{U}) - \textbf{U} \textbf{w}^T = \textbf{F}^\prime (\textbf{U}^\prime) \\
      &+ 
      \left( \begin{array}{c} 
        0 \\
        \rho \textbf{w} \left( \textbf{v} - \textbf{w} \right)^T \\
        \rho \left( \textbf{v} \textbf{w} \right) \left( \textbf{v} - \textbf{w} \right) - \frac{\rho}{2} \textbf{w}^2 \left( \textbf{v} - \textbf{w} \right) 
        + p \textbf{w} - \textbf{B} \left( \textbf{w} \cdot \textbf{B} \right) \\
        - \textbf{w} \textbf{B}^T \\
        0
      \end{array} \right)
    \end{aligned} \end{equation}
    Finally, we can resubstitute the fluxes obtained from solving the Riemann problem in the moving frame into the additional terms:
    \begin{equation} \begin{aligned}
      \textbf{F}_m(\textbf{U}) &= \textbf{F}_s(\textbf{U}) - \textbf{U} \textbf{w}^T =  \\
      &= \textbf{F}^\prime (\textbf{U}^\prime) + 
      \left( \begin{array}{c} 
        0 \\
        \textbf{w} \textbf{Q}_1^T \\
        \textbf{w} \textbf{Q}_2 + \frac{1}{2} \textbf{Q}_1 \textbf{w}^2 \\
        - \textbf{w} \textbf{B}^T \\
        0
      \end{array} \right).
    \end{aligned} \end{equation}
With this procedure, the discretized MHD equations always retain their
upwind character, and a stable evolution on a dynamic Voronoi mesh is obtained.

  \section{Test problems}

  In this section we study a number of standard test problems that are
  essential for validating the accuracy of a numerical MHD
  implementation.  We consider first simple magnetic shock tubes and
  then move to more demanding two-dimensional problems, each designed
  to test different aspects of the code.
 
    \label{sec:tests}

    \subsection{Shocktube}

    The results of a standard one-dimensional shock-tube problem with
    the initial conditions described by \citet{keppens2004a} are shown in
    Fig.~\ref{fig:shock}. This shock-tube problem has the advantage that all
    seven MHD waves are present. For comparison it is run for static and
    moving grids with two different Riemann solvers each (HLLD and Lax-Friedrich).
    Overall, all runs show good agreement. Upon close inspection, however, there
    are some differences. As expected, the HLLD-solver is less diffusive than
    the Lax-Friedrich solver. In addition, the moving grid behaves particularily
    better at the slow shock and the contact discontinuity, similar to previous
    studies \citep{vandam2006a}.
    
    We note that in contrast to this one-dimensional shock-tube, all
    multi-dimensional problems we discuss next are run with an activated
    divergence cleaning scheme. Also, they all apply periodic boundary
    conditions and use $\gamma = {5}/{3}$ for the gas.

    \begin{figure}
      \centering
      \includegraphics{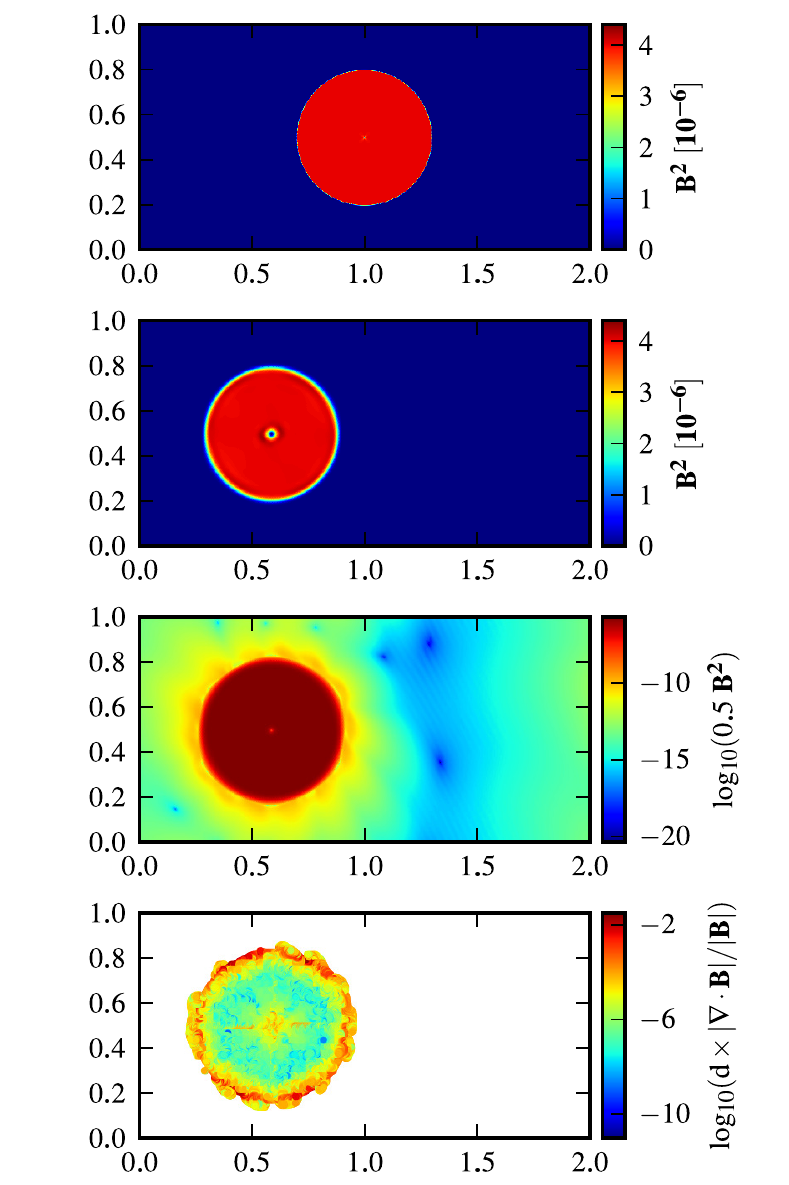}
      \caption[Two-dimensional Loop advection]
      {Magnetic energy evolution of a field loop advected over a
        moving grid with a resolution of $1600 \times 800$ cells. The
        two panels on top show the magnetic energy density at times
        $t=0$ (first row) and $t=18$ (second row) on a linear color
        scale, while the third panel repeats the $t=18$ result with a
        logarithmic color scale. Finally, the bottom panel shows the
        relative importance of the divergence error of the magnetic
        field by comparing it to the magnetic field itself for all
        cells with $|\textbf{B}| > 10^{-5}$. Here, $d$ is the approximative
        size of a cell calculated from its volume assuming that it is spherical.}
      \label{fig:loop}
    \end{figure}

    \subsection{Advection of a magnetic loop}
 
    In this test problem, a magnetic field loop is advected by a
    constant velocity field. The magnetic field itself is too small to
    be dynamically important.  The initial conditions are defined in a
    box of extension $0 < x < 2$, $0 < y < 1$ and are given by
    \begin{equation} \begin{aligned}
      \rho &= 1 \\
      p &= 1 \\
      \textbf{v} & = \left( \sin( \pi / 3 ), \cos( \pi / 3 ), 0 \right) \\
      \textbf{A} &= \left( 0, 0, \max( 0.001 \times (0.3 - r), 0 ) \right)
    \end{aligned} \end{equation} where $r$ is the radial distance to
  the center of the loop and $\textbf{A}$ the vector potential. The
  initial magnetic field is directly calculated from the vector
  potential. We use a hexagonal grid with $1600\times 800$ cells in
  total.

  In Figure~\ref{fig:loop}, we show the magnetic field strength and
  its divergence error after the loop has crossed the box several
  times. The loop is preserved extremely well, and advection errors
  are very small, which highlights the principal advantage of the
  moving mesh code. Notice that the logarithmic plot of the magnetic
  energy density in Fig.~\ref{fig:loop} demonstrates that the smearing
  at the edge of the loop is quite small. More importantly, the
  relative error of the divergence of the magnetic field is smaller
  than $\sim 10^{-3}$ for all parts of the magnetic loop except the very
  outer edge where the magnetic field drops to zero. The
  error at the center of the loop is caused by insufficient angular
  resolution there.
    
     \begin{figure}
      \centering
      \includegraphics{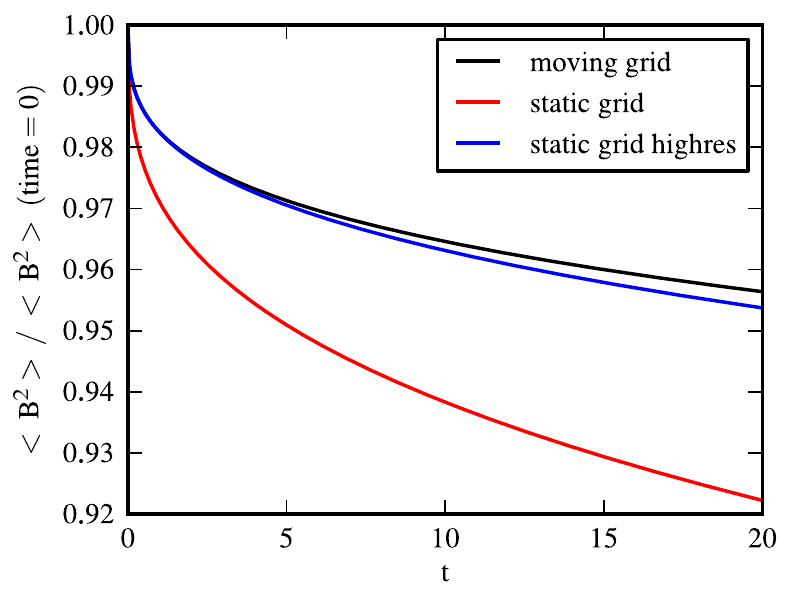}
      \caption[Magnetic energy evolution in Loop advection problem]
      {Time evolution of the magnetic energy in an advected field loop. The
        change of the magnetic energy is shown for a moving grid of
        $800 \times 400$ cells (black), a static grid of $800 \times
        400$ cells (red) and a static grid of $1600 \times 800$ cells
        (blue). }
      \label{fig:loopenergy}
    \end{figure}
    
    For a quantitative check of the advection properties of our MHD
    code we compare in Fig.~\ref{fig:loopenergy} the time evolution of
    the magnetic energy with simulations on a static grid with the
    same resolution and a higher resolution by a factor of two. The
    moving mesh implementation conserves the magnetic energy
    significantly better than a simulation on a static grid of the
    same resolution. In fact, it is even a little bit better than a
    simulation on a static grid with a twice better spatial
    resolution.
    
    \begin{figure}
      \centering
      \includegraphics[width=\linewidth]{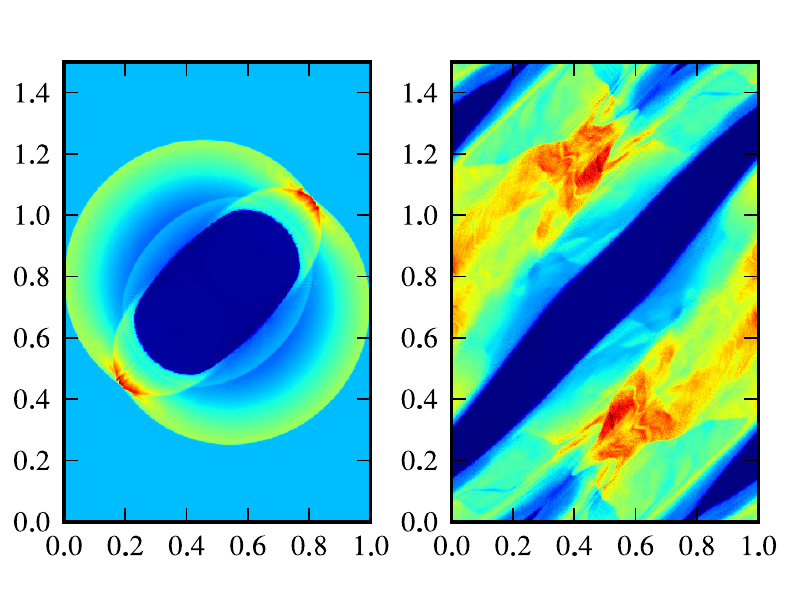}
      \caption[Two-dimensional spherical blast wave]
      {Linear density maps of a two-dimensional spherical blast wave
        at times $t=0.2$ (left) and $t=3.75$ (right). The density
        mapped to the color scale ranges
        from $0.1$ to $3.0$ (left) and $0.1$ to $2.4$ (right). The
        system is resolved with $512 \times 768$ cells.}
      \label{fig:blast}
    \end{figure}
    
    \subsection{Magnetic blast wave}

    The magnetic blast wave problem consists of an initially circular,
    overpressurized region in a magnetized fluid. The initial conditions
    are given by
    \begin{equation} \begin{aligned}
      \rho &= 1 \\
      p &=  \left\{ \begin{array}{ll} &10 \ r < 0.1 \\  &0.1 \ r \leq 0.1 \end{array} \right. \\
      \textbf{v}  &= \left( 0, 0, 0 \right) \\
      \textbf{B}  &= \left(  \frac{1}{\sqrt{2}}, \frac{1}{\sqrt{2}}, 0 \right)
    \end{aligned} \end{equation} in a region of size $0 < x < 1$, $0 <
  y < 1.5$.  Fig.~\ref{fig:blast} shows the density soon after the
  start of a simulation with a resolution of $512 \times 768$ cells,
  as well at a later time when the shock has already interacted with
  itself after crossing the box. The early snapshot shows how the
  shock wave becomes anisotropic as the magnetic field introduces a
  preferred direction.

  The code captures the complex shock dynamics very well and manages
  to resolve quite sharp interfaces, such that the results favourably
  compare to corresponding calculations with other MHD codes
  \citep[see, e.g.][]{Fromang2006,stone2008a,Rosswog2007}. Note that
  there are some corrugations in the red regions behind the shock front
  in left panel of Fig.~\ref{fig:blast} that originate from small distortions
  in the grid. This can be cured by adding a refinement scheme which
  splits elongated cells.
    
    \begin{figure*}
      \centering
      \includegraphics{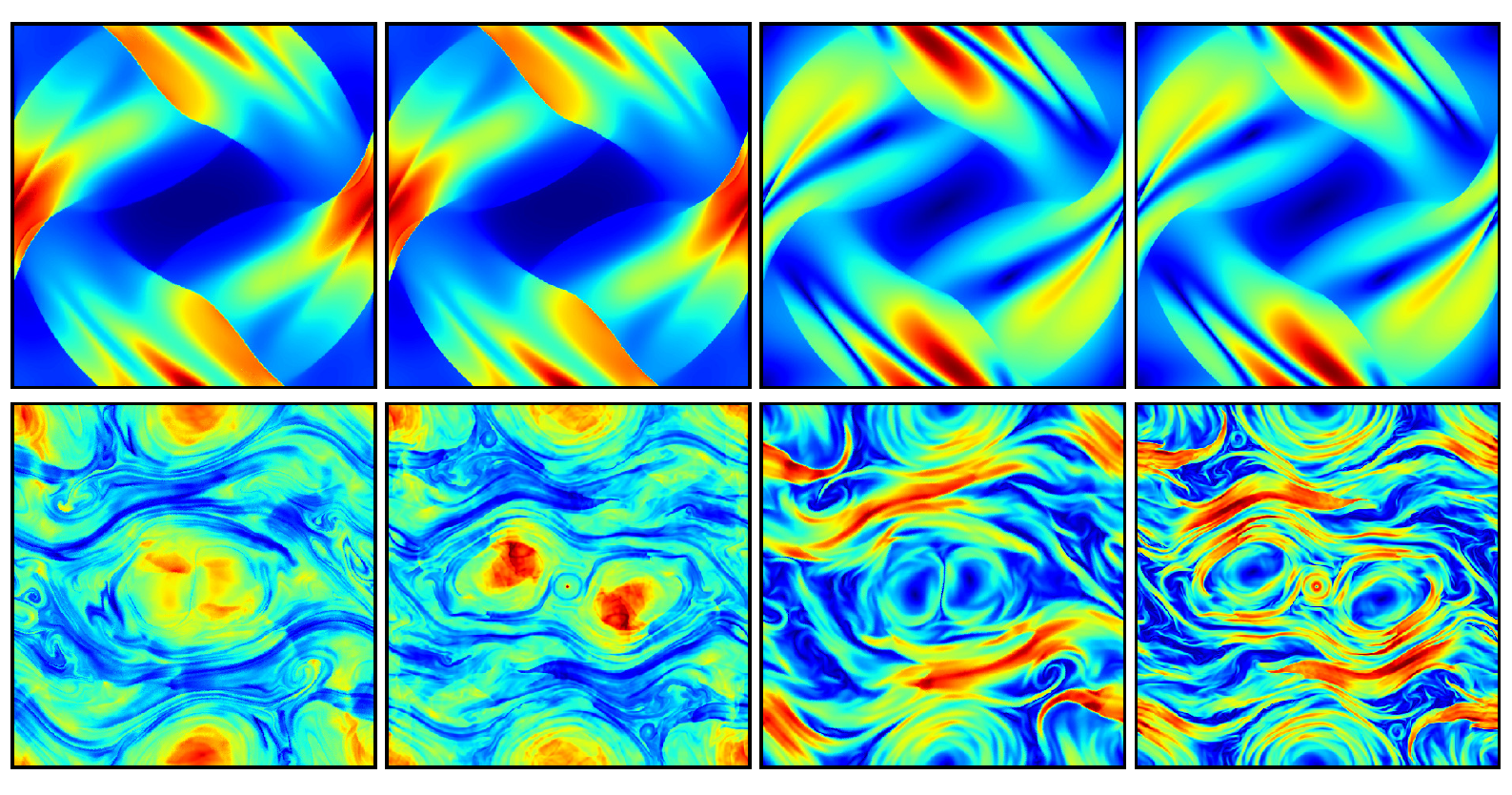}
      \caption[Two-dimensional Orszag-Tang vortex]
      {Two-dimensional Orszag-Tang vortex test. Shown are linear
        density maps (pairs on the left) and magnetic energy density
        maps (pairs on the right) at times $t=0.25$ (left) and $t=2.5$
        (right). For each pair, the left image shows the result of the
        {\small AREPO} code, while the right images displays the same
        simulation when the {\small ATHENA} code \citep{stone2008a} is
        used. The number of resolution elements is identical and set
        to $600 \times 600$ cells on both cases. The density ranges
        displayed in the color maps cover $0.06$ to $0.5$ (top left
        pair) and $0.1$ to $0.4$ (top right pair), respectively, and
        for the magnetic energy $0.0$ to $0.4$ (bottom left pair) and
        $0.0$ to $0.3$ (bottom right pair), respectively.}
      \label{fig:vortex}
    \end{figure*}
    
    \subsection{Orszag-Tang vortex}

\begin{figure}
      \centering
      \includegraphics{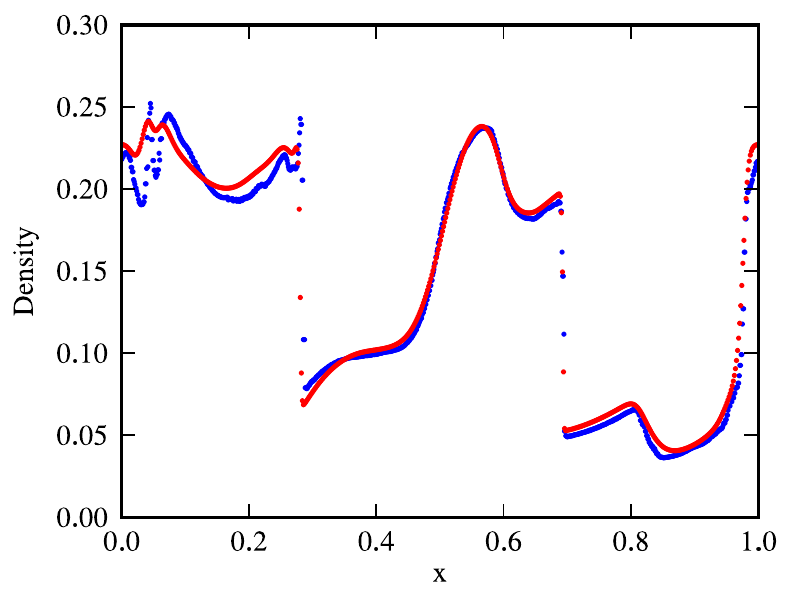}
      \caption[Slice through the Orszag-Tang vortex]
      {Slice of the density at $y=0.33$ and $t=0.5$ of the Orszag-Tang
      vortex. Blue and red symbols show the results for the {\small AREPO}
      and the {\small ATHENA} code.}
      \label{fig:vortex2}
    \end{figure}

    The so-called Orszag-Tang vortex \citep{orszag1979a} is an
    often employed two-dimensional test for MHD codes. Besides
    being an excellent stability test it also examines how shocks 
    interact with each other and produce  a supersonically
    turbulent system where the turbulence decays with time.
    The initial conditions as first described by \citet{picone1991a}
    are defined by
    \begin{equation} \begin{aligned}
      \rho &= \frac{ \gamma^2  }{  4 \pi } \\
      p &=  \frac{ \gamma }{  4 \pi }\\
      \textbf{v}  &= \left( -\sin( 2 \pi y ), \sin( 2 \pi x ), 0 \right) \\
      \textbf{B}  &= \left( -\sin( 2 \pi y ), \sin( 4 \pi x ), 0 \right)
    \end{aligned} \end{equation} in a box of extension $0 < x$, $y <
  1$. We consider test runs with a resolution of $600 \times 600$
  moving mesh cells. Fig.~\ref{fig:vortex} compares the density and
  magnetic energy in our simulation with results obtained with the
  {\small ATHENA} code \citep{stone2008a}, at two different times. In
  the earlier snapshot several shocks are just building up, while at
  the time of the second snapshot the simulation has already evolved
  for some time and become fully turbulent. A quantitative comparison
  at $t=0.5$ is shown in Fig.~\ref{fig:vortex2}. The agreement between the
  two different codes is in general very good and reassuring. Note
  that in contrast to the {\small ATHENA} code the magnetic version of
  {\small AREPO} does not keep the initial symmetry perfectly.  This
  deviation from perfect symmetry is in part a result of the order in
  which the fluxes over different interfaces of a cell are applied. As
  this is more or less random in {\small AREPO}, small round-off
  asymmetries are introduced even for perfectly symmetric conditions,
  and those tend to be amplified by means of the additional freedom
  allowed by the moving mesh.  In addition, the {\small ATHENA} code
  shows slightly finer structures at later times.
  
  Due to the different underlying schemes, Voronoi grid in {\small AREPO}
  and Cartesian grid in {\small ATHENA}, there are also differences in
  the computational resources the two codes require. In the Orszag-Tang
  vortex problem the {\small AREPO} consumes about three times the 
  amount of memory the {\small ATHENA} code needs. This is mainly
  associated with the construction of the Voronoi grid which has to be
  done in each timestep. In addition, {\small ATHENA} is up to six
  times faster than {\small AREPO}. This is caused by a combination
  of two effects. The timesteps in the {\small AREPO} code are smaller
  by a factor of about three due to stricter timestep criterions \footnote{
  Using a less conservative timestep criterion can eliminate this contribution
  to the difference}, and a 
  single timestep takes twice as long for {\small AREPO} than for
  {\small ATHENA} on average. The higher timestep cost is mainly caused
  by the construction of the Voronoi grid which takes about $40\%$
  of all CPU time, and the gradient and flux calculations which are slowed
  down considerably by cache misses in {\small AREPO} because the grid
  cells are stored unordered in memory. In addition, {\small AREPO} needs
  to solve more Riemann problems per cell. We note, however, that this difference
  is particularly large in this application since the problem is very simple and dominated
  by magnetohydrodynamics only. For more complex applications that
  are often dominated by the evaluation of self-gravity this difference becomes
  much less important.

  \section{Driven supersonic MHD turbulence}

    \label{sec:turbulence}

    Arguably one of the most challenging test problems for an MHD-code
    is driven supersonic MHD-turbulence, which we consider in this
    section. To this end, we create a periodic box of unit size
    containing an isothermal gas with constant density. The initially
    uniform magnetic field has a strength of $B=0.1\sqrt{3}$. In our
    dimensionless system of units, the volume, density and speed of
    sound are given by $V = \rho = c_s = 1$, respectively. To drive
    the turbulence, an external force field is applied to the fluid
    which we implement as in \citet{price2010a}.  The force field is
    setup in Fourier space and only contains power in a small range of
    low frequency modes between $k_{\rm min}$ and $k_{\rm max}$. The
    amplitude of the force field is given by a paraboloid with a maximum
    at  $k_c = (k_{\rm max}-k_{\rm min})/2$.
    The individual phases of the modes are drawn from
    Ornstein-Uhlenbeck processes. The stirring force field is then
    obtained grid by an inverse Fourier transformation at each cell
    center.

\begin{figure}
      \centering
      \includegraphics{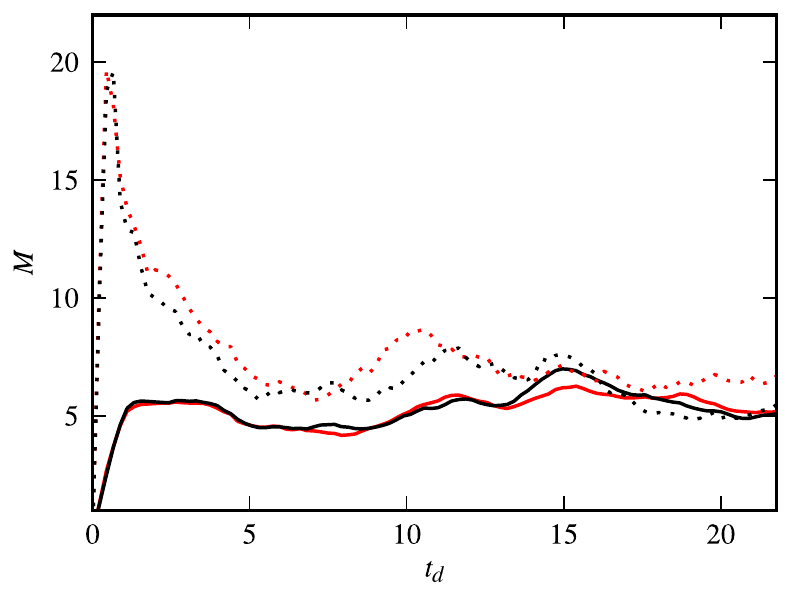}
      \caption[rms Mach number]
      {The solid (dashed) lines show the sonic (Alfv\'enic) Mach
        number as a function of time. Both are shown for two
        simulations with a resolution of $64^3$ cells (red) and
        $128^3$ cells (black).}
      \label{fig:mach}
    \end{figure}

    The amplitude of the power spectrum of the driving forces is
    chosen to lead to an {\em rms} Mach number of about $M_S\simeq 5.5$ once
    stationary turbulence has built up. The time evolution of this
    quantity, together with that of the approximate Alfv\'enic Mach
    number $M_A = \sqrt{2\langle \rho u^2\rangle / \langle
      \textbf{B}^2\rangle}$ is shown in Figure~\ref{fig:mach}. At the
    beginning, the Alfv\'enic Mach number increases rapidly, because
    the fluid velocity increases faster than the magnetic field. Later
    it decreases again as the magnetic field catches up.  We note that
    the sonic mach numbers of our two simulations with a resolution of
    $64^3$ and $128^3$ cells agree well with each other for an
    identical driving, as expected. However, the Alfv\'enic Mach
    number of the high resolution run is slightly smaller on average.  The
    expected dynamical timescale in our setup is $t_d = L/(2 M_S)$,
    and we find that after about three dynamical time scales the
    turbulence is fully established. For the following analysis we
    therefore only consider outputs after $t=4\,t_d$.

    To calculate power spectra of the turbulent velocity fields, we
    interpolate the Voronoi cells to a uniform Cartesian grid using
    the Voronoi interpolation, or in other words, we sample the
    reconstructed velocity field defined on the Voronoi mesh at the
    coordinates of a fine Cartesian grid.  To compare kinetic and
    magnetic power spectra, we also calculate the kinetic power
    spectra for a density-weighted velocity field, $\textbf{u} =
    \sqrt{\rho} \textbf{v}$.  The magnetic power spectrum is obtained
    directly from the magnetic field. In this way, the integral over
    the power spectra is equivalent to the total kinetic and magnetic
    energy, respectively.

\begin{figure}
      \centering
      \includegraphics{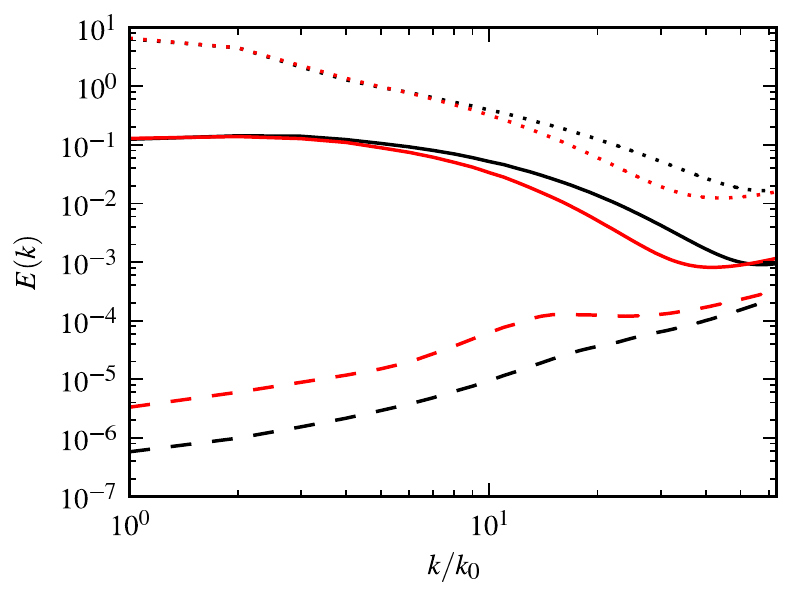}
      \caption[power spectrum]
      {Power spectra of driven MHD turbulence. Shown are the total
        magnetic power spectrum (solid lines), the divergence part of
        the magnetic power spectrum (dashed lines) and the velocity
        power spectrum (dotted lines). Spectra are shown for two runs
        with a resolution of $64^3$ cells (red) and $128^3$ cells
        (black), respectively.  }
      \label{fig:turbulence}
    \end{figure}

    The resulting magnetic and velocity power spectra are shown in
    Figure~\ref{fig:turbulence}. They are averaged over $70$ snapshots
    ranging from $t=4.4\,t_d$ to $22.0\,t_d$.  The velocity spectra at
    low wave numbers are largely dominated by the driving
    mechanism. Towards smaller scales a power-law for the kinetic
    energy is found, followed by a dissipative regime when the Nyquist
    frequency is approached. The inertial range of the turbulence is
    however quite small at this comparatively low resolution. This
    probably also explains why no clear power law region for the
    magnetic power spectrum is seen. Note that in this setup the
    magnetic energy is about $1.5$ orders of magnitude lower than the
    kinetic energy.  The shape of the power spectra is similar to those
    found by \citet{kritsuk2011a} in a systematic comparison of
    different MHD codes, suggesting that our MHD implementation is
    consistent with the results found there. This is particularly
    encouraging as we here use a fully dynamic mesh, for which it is
    quite demanding to control $\nabla\cdot \textbf{B}$ errors
    under conditions of highly supersonic isothermal turbulence.

    We examine the latter point quantitatively by measuring the
    divergence part of the magnetic power spectra, obtained through
    the use of a Helmholtz decomposition in Fourier space.  In the
    higher resolution run ($128^3$), the $\nabla\cdot\textbf{B}$ error is
    smaller by roughly an order of magnitude compared with the low
    resolution run ($64^3$), and in both cases, the error is very
    small compared to the actual strength of the magnetic field.

  \section{Collapse of a magnetized cloud}
    \label{sec:collapse}
    
    \begin{figure*}
      \centering
      \includegraphics{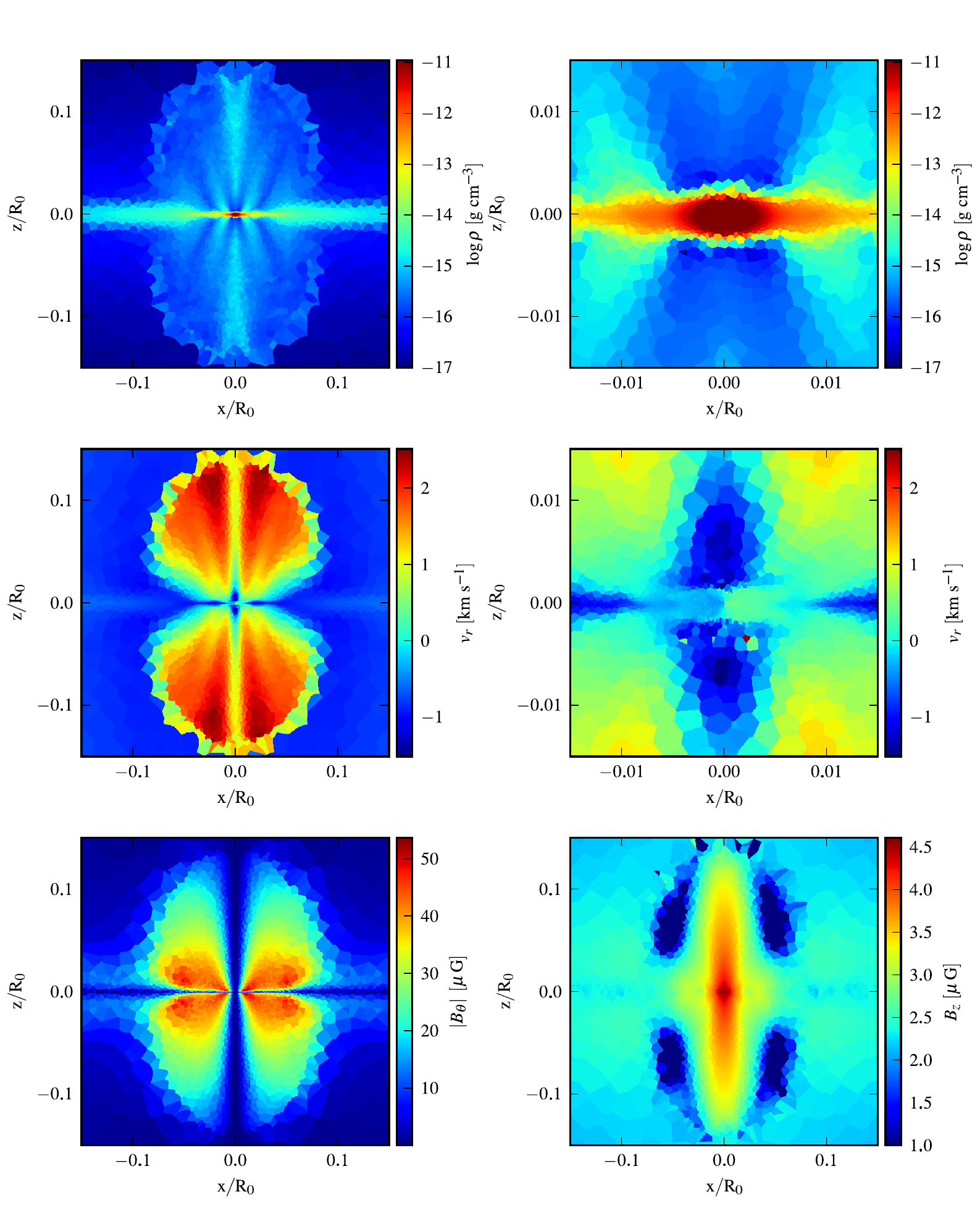} 
      \caption[Collapse]
      {The collapsed cloud after $1.13\, \mathrm{t_{ff}}$. Shown are the density
      (top row), inflow/outflow velocity (middle row),
      magnetic field in the $z$-direction (lower left plot) and in the azimuthal direction
      (lower right plot), respectively.}
      \label{fig:collapse}
    \end{figure*}

    Finally, in this section we apply the code to the collapse of a
    magnetized sphere, which can be viewed as an idealized
    astrophysical problem that is of direct relevance to star
    formation.  We choose initial conditions similar to the study of
    \citet{hennebelle2008a}. A rigidly rotating homogenous sphere with a 
    radius of $R_0 = 0.015\,\mathrm{pc}$ and a total mass of one
    solar mass is embedded in an atmosphere that is $100$ times less dense, 
    with a small transition region at the boundary. With an initial density of
    $4.8 \times 10^{-18} \mathrm{g\ cm^{-3}}$ the freefall time is $3 \times 10^4$
    years. The sphere rotates rigidly with a period of $4.7 \times
    10^5$ years, equivalent to a ratio of rotational over
    gravitational energy equal to $0.045$. The whole box in which the
    cloud is embedded is filled with a uniform magnetic field parallel
    to the rotation axis with a strength of $3.0\,\mathrm{\mu G}$,
    equivalent to a mass-to-flux over critical mass-to-flux ratio of
    $20$. We adopt the same equation of state as
    \citet{hennebelle2008a} given by
    \begin{equation}
      P = \rho \times \left( c_s^0 \right)^2 \times \left[1 + (\rho/\rho_c)^{4/3} \right]^{0.5},
    \end{equation}
    with $c_s^0 = 0.2\, \mathrm{km\ s^{-1}}$ and $\rho_c = 10^{-13}\,\mathrm{g\ cm^{-3}}$.
    Our simulation box has a size of $0.06 \mathrm{pc}$ and inflow/outflow
    boundary conditions with an initial resolution of $128^3$ cells. We apply a
    refinement criterion that splits a cell when its freefall timescale becomes
    smaller than ten times its sound crossing timescale. However, we limit the
    volume of a cell to be not smaller than $5 \times 10^{-17} \mathrm{pc}$ which
    is equivalent to an effective resolution of $16384^3$ cells.
    
    After $1.13 t_\mathrm{ff}$ the number of cells in the simulation increased
    from $\approx 2.1$ million in the beginning to $\approx 2.2$ million. In contrast to the
    initial setup, however, most of the cells are clustered in a very small region at
    the center of the cloud where most of the mass resides. The cloud at this time is shown
    in Fig.~\ref{fig:collapse}. As expected, a proto star has formed surrounded by an
    accretion disk with a radius of about $0.03\,R_0$. Magnetically powered outflows are launched
    along the $z$-axis with a velocity of $2.\, \mathrm{km\ s^{-1}}$. The magnetic field
    parallel to the $z$-axis has been amplified by compression beyond $10^4\,\mathrm{\mu G}$
    close to the protostar. An azimuthal magnetic field has been generated during the
    collapse with a typical strength of $5\, \mathrm{\mu G}$, up to $10^5\, \mathrm{\mu G}$
    at the protostar, in good qualitative agreement with \citet{hennebelle2008a}.

    Fig.~\ref{fig:collapse} also nicely shows how the adaptive mesh adapts to the
    internal structure of the fluid in the simulation and is able to handle large density
    contrasts very well.
    
  \section{Conclusions}
    \label{sec:conclusions}

    In this study, we have presented our implementation of ideal MHD
    in the moving-mesh code {\small AREPO}. The numerical scheme
    employs a fully adaptive Voronoi mesh which can freely move with
    the flow. The dynamics is solved with a second-order accurate
    finite-volume scheme that employs a spatial reconstruction and a
    flux calculation at each mesh interface, based on approximate
    solutions of Riemann problems.  To maintain the divergence
    constraint of the magnetic field we have implemented the
    divergence cleaning scheme proposed by \citet{dedner2002a}. In
    contrast to the constrained transport approach, this method can be
    readily implemented for unstructured dynamic meshes that we use here.

    To our knowledge, our new MHD implementation is the first
    three-dimensional Lagrangian mesh code capable of following
    magnetic fields. There already exist SPH implementations of
    Lagrangian MHD \citep[e.g.][]{Rosswog2007,Dolag2009}, but they still
    suffer quite a bit from the inherent subsonic noise in SPH,
    necessitating relatively aggressive cleaning schemes, and from the
    relatively slow convergence rate of SPH \citep{Springel2010c}. Our new scheme fairs much
    better in this respect, retaining the high accuracy for shocks and
    smooth flows that is reached by ordinary fixed-mesh codes. In
    addition, our method drastically reduces advection errors in cases where
    large bulk velocities occur. Especially in this regime, our
    magnetic version of {\small AREPO} can be expected to surpass the accuracy of
    corresponding fixed mesh codes.

    We have verified the code's performance in a number of
    standard test problems, ranging from simple magnetic shock tubes
    to complicated interactions of multiple shock waves such as the
    Orszag-Tang vortex problem. In all our tests we found satisfactory
    agreement with published results from fixed-mesh MHD codes.  We
    also applied our code to the collapse of a magnetized cloud of one
    solar mass under self-gravity, nicely reproducing results obtained
    by \citet{hennebelle2008a} with the {\small RAMSES} code.  In
    addition, we have used our new code to simulate driven, highly
    supersonic isothermal magnetic turbulence in a periodic box. This
    has established that the new moving-mesh MHD code can handle this
    demanding problem robustly.  The resulting power spectra are in
    good agreement to previous work considering that we have used a coarser
    resolution than in recent dedicated MHD turbulence studies. In future work,
    it will be interesting to study the details of the statistics of
    the turbulence represented by the {\small AREPO} code, but this is
    beyond the scope of this paper and requires much larger production
    simulations.
    
    We conclude that our new MHD version of {\small AREPO} can be
    applied with confidence to a large range of astrophysical
    problems. Its Lagrangian properties should make it particularly
    useful for studies of star formation, magnetic fields in galaxies
    and clusters of galaxies, and accretion disks.  We also note that
    the code includes a powerful and accurate gravity solver, as well
    as modules for collisionless dynamics, making it also attractive
    for simulations of cosmic structure formation.

  \section{Acknowledgements}
  We want to thank James M. Stone for making the {\small ATHENA} code
  publicly available, and Daniel Price and Christoph Federrath for
  providing us with a version of their turbulence driving routines. We
  further thank Tom Abel, Ewald M\"uller, Martin Obergaullinger,
  Fabian Miczek and the anonymous referee for insightful discussions.  R.~P. gratefully
  acknowledges financial support of the Klaus Tschira Foundation.

  \bibliographystyle{mn2e}

\begin{thebibliography}{}

\bibitem[\protect\citeauthoryear{{Balbus} \& {Hawley}}{{Balbus} \&
  {Hawley}}{1998}]{Balbus1998}
{Balbus} S.~A.,  {Hawley} J.~F.,  1998, Reviews of Modern Physics, 70, 1

\bibitem[\protect\citeauthoryear{{Brackbill} \& {Barnes}}{{Brackbill} \&
  {Barnes}}{1980}]{Brackbill1980}
{Brackbill} J.~U.,  {Barnes} D.~C.,  1980, Journal of Computational Physics,
  35, 426

\bibitem[\protect\citeauthoryear{{Brandenburg}}{{Brandenburg}}{2010}]{Brandenburg2010}
{Brandenburg} A.,  2010, \mnras, 401, 347

\bibitem[\protect\citeauthoryear{{Brandenburg} \& {Dobler}}{{Brandenburg} \&
  {Dobler}}{2002}]{Brandenburg2002}
{Brandenburg} A.,  {Dobler} W.,  2002, Computer Physics Communications, 147,
  471

\bibitem[\protect\citeauthoryear{{Collins}, {Xu}, {Norman}, {Li} \&
  {Li}}{{Collins} et~al.}{2010}]{Collins2010}
{Collins} D.~C.,  {Xu} H.,  {Norman} M.~L.,  {Li} H.,    {Li} S.,  2010, \apjs,
  186, 308

\bibitem[\protect\citeauthoryear{Darwish \& Moukalled}{Darwish \&
  Moukalled}{2003}]{darwish2003a}
Darwish M.~S.,  Moukalled F.,  2003, International Journal of Heat and Mass
  Transfer, 46, 599

\bibitem[\protect\citeauthoryear{{Dedner}, {Kemm}, {Kr{\"o}ner}, {Munz},
  {Schnitzer} \& {Wesenberg}}{{Dedner} et~al.}{2002}]{dedner2002a}
{Dedner} A.,  {Kemm} F.,  {Kr{\"o}ner} D.,  {Munz} C.,  {Schnitzer} T.,
  {Wesenberg} M.,  2002, Journal of Computational Physics, 175, 645

\bibitem[\protect\citeauthoryear{{Dolag} \& {Stasyszyn}}{{Dolag} \&
  {Stasyszyn}}{2009}]{Dolag2009}
{Dolag} K.,  {Stasyszyn} F.,  2009, \mnras, 398, 1678

\bibitem[\protect\citeauthoryear{{Duffell} \& {MacFadyen}}{{Duffell} \&
  {MacFadyen}}{2011}]{Duffell2011}
{Duffell} P.~C.,  {MacFadyen} A.~I.,  2011, ArXiv e-prints, 1104.3562

\bibitem[\protect\citeauthoryear{{Evans} \& {Hawley}}{{Evans} \&
  {Hawley}}{1988}]{evans1988a}
{Evans} C.~R.,  {Hawley} J.~F.,  1988, \apj, 332, 659

\bibitem[\protect\citeauthoryear{{Fromang}, {Hennebelle} \&
  {Teyssier}}{{Fromang} et~al.}{2006}]{Fromang2006}
{Fromang} S.,  {Hennebelle} P.,    {Teyssier} R.,  2006, \aap, 457, 371

\bibitem[\protect\citeauthoryear{{Gardiner} \& {Stone}}{{Gardiner} \&
  {Stone}}{2005}]{gardiner2005a}
{Gardiner} T.~A.,  {Stone} J.~M.,  2005, Journal of Computational Physics, 205,
  509

\bibitem[\protect\citeauthoryear{Harten, Lax \& Van~Leer}{Harten
  et~al.}{1983}]{harten1983a}
Harten A.,  Lax P.~D.,    Van~Leer B.,  1983, SIAM Review, 25, 35

\bibitem[\protect\citeauthoryear{{Hennebelle} \& {Fromang}}{{Hennebelle} \&
  {Fromang}}{2008}]{hennebelle2008a}
{Hennebelle} P.,  {Fromang} S.,  2008, \aap, 477, 9

\bibitem[\protect\citeauthoryear{{Keppens}}{{Keppens}}{2004}]{keppens2004a}
{Keppens} R.,  2004, Fusion Science and Technology, 45, 2, 107

\bibitem[\protect\citeauthoryear{{Keppens}, {Nool}, {T{\'o}th} \&
  {Goedbloed}}{{Keppens} et~al.}{2003}]{keppens2003a}
{Keppens} R.,  {Nool} M.,  {T{\'o}th} G.,    {Goedbloed} J.~P.,  2003, Computer
  Physics Communications, 153, 317

\bibitem[\protect\citeauthoryear{{Kritsuk}, {Nordlund}, {Collins}, {Padoan},
  {Norman}, {Abel}, {Banerjee}, {Federrath}, {Flock}, {Lee}, {Li}, {Mueller},
  {Teyssier}, {Ustyugov}, {Vogel} \& {Xu}}{{Kritsuk}
  et~al.}{2011}]{kritsuk2011a}
{Kritsuk} A.~G.,  {Nordlund} A.,  {Collins} D.,  {Padoan} P.,  {Norman} M.~L.,
  {Abel} T.,  {Banerjee} R.,  {Federrath} C.,  {Flock} M.,  {Lee} D.,  {Li}
  P.~S.,  {Mueller} W.-C.,  {Teyssier} R.,  {Ustyugov} S.~D.,  {Vogel} C.,
  {Xu} H.,  2011, ArXiv e-prints, 1103.5525

\bibitem[\protect\citeauthoryear{{Miyoshi} \& {Kusano}}{{Miyoshi} \&
  {Kusano}}{2005}]{miyoshi2005a}
{Miyoshi} T.,  {Kusano} K.,  2005, Journal of Computational Physics, 208, 315

\bibitem[\protect\citeauthoryear{{Orszag} \& {Tang}}{{Orszag} \&
  {Tang}}{1979}]{orszag1979a}
{Orszag} S.~A.,  {Tang} C.-M.,  1979, Journal of Fluid Mechanics, 90, 129

\bibitem[\protect\citeauthoryear{{Picone} \& {Dahlburg}}{{Picone} \&
  {Dahlburg}}{1991}]{picone1991a}
{Picone} J.~M.,  {Dahlburg} R.~B.,  1991, Physics of Fluids B, 3, 29

\bibitem[\protect\citeauthoryear{{Powell}, {Roe}, {Linde}, {Gombosi} \& {de
  Zeeuw}}{{Powell} et~al.}{1999}]{powell1999a}
{Powell} K.~G.,  {Roe} P.~L.,  {Linde} T.~J.,  {Gombosi} T.~I.,    {de Zeeuw}
  D.~L.,  1999, Journal of Computational Physics, 154, 284

\bibitem[\protect\citeauthoryear{{Price}}{{Price}}{2010}]{Price2010b}
{Price} D.~J.,  2010, \mnras, 401, 1475

\bibitem[\protect\citeauthoryear{{Price} \& {Federrath}}{{Price} \&
  {Federrath}}{2010}]{price2010a}
{Price} D.~J.,  {Federrath} C.,  2010, \mnras, 406, 1659

\bibitem[\protect\citeauthoryear{{Rosswog} \& {Price}}{{Rosswog} \&
  {Price}}{2007}]{Rosswog2007}
{Rosswog} S.,  {Price} D.,  2007, \mnras, 379, 915

\bibitem[\protect\citeauthoryear{Rusanov}{Rusanov}{1961}]{rusanov1961a}
Rusanov V.~V.,  1961, J. Comput. Math. Phys. USSR, 1, 267–279

\bibitem[\protect\citeauthoryear{{Springel}}{{Springel}}{2010a}]{springel2010a}
{Springel} V.,  2010a, \mnras, 401, 791

\bibitem[\protect\citeauthoryear{{Springel}}{{Springel}}{2010b}]{Springel2010c}
{Springel} V.,  2010b, \araa, 48, 391

\bibitem[\protect\citeauthoryear{{Stone}, {Gardiner}, {Teuben}, {Hawley} \&
  {Simon}}{{Stone} et~al.}{2008}]{stone2008a}
{Stone} J.~M.,  {Gardiner} T.~A.,  {Teuben} P.,  {Hawley} J.~F.,    {Simon}
  J.~B.,  2008, \apjs, 178, 137

\bibitem[\protect\citeauthoryear{{Stone} \& {Norman}}{{Stone} \&
  {Norman}}{1992}]{Stone1992}
{Stone} J.~M.,  {Norman} M.~L.,  1992, \apjs, 80, 791

\bibitem[\protect\citeauthoryear{{Toro}}{{Toro}}{1997}]{Toro1997}
{Toro} E.,  1997, Riemann solvers and numerical methods for fluid dynamics.
Springer

\bibitem[\protect\citeauthoryear{{T{\'o}th}}{{T{\'o}th}}{2000}]{toth2000a}
{T{\'o}th} G.,  2000, Journal of Computational Physics, 161, 605

\bibitem[\protect\citeauthoryear{{van Dam} \& {Zegeling}}{{van Dam} \&
  {Zegeling}}{2006}]{vandam2006a}
{van Dam} A.,  {Zegeling} P.~A.,  2006, Journal of Computational Physics, 216,
  526

\bibitem[\protect\citeauthoryear{{van Leer}}{{van Leer}}{1974}]{vanleer1974a}
{van Leer} B.,  1974, Journal of Computational Physics, 14, 361

\bibitem[\protect\citeauthoryear{{van Leer}}{{van Leer}}{1984}]{Leer1984}
{van Leer} B.,  1984, SIAM J. Sci. Stat. Comput., 5, 1

\bibitem[\protect\citeauthoryear{{Ziegler} \& {Yorke}}{{Ziegler} \&
  {Yorke}}{1997}]{Ziegler1997}
{Ziegler} U.,  {Yorke} H.~W.,  1997, Computer Physics Communications, 101, 54

\end{thebibliography}

  \label{lastpage}

\end{document}